\documentclass{aa}
\usepackage{graphicx}
\newcommand\msun{\rm\,M_\odot}

\usepackage{txfonts}
\usepackage{natbib}
\usepackage{gensymb}

\begin{document}

  \title{Ionization processes in a local analogue of distant clumpy galaxies: VLT MUSE IFU spectroscopy and FORS deep images\thanks{Based on observations collected at the European Organisation for Astronomical Research in the Southern Hemisphere, Chile: ESO MUSE program 60.A-9320(A) and FORS program 382.B-0213(A)} of the TDG NGC~5291N}
\author{
 J. Fensch
         \inst{1}
          \and
          P.-A. Duc \inst{1}
          \and
          P.~M. Weilbacher \inst{2}
          \and
          M. Boquien \inst{3,4}
          \and
          E. Zackrisson \inst{5}
          }

   \offprints{J. Fensch (jeremy.fensch@cea.fr)}

   \institute{
              Laboratoire AIM Paris-Saclay, CEA/IRFU/SAp, Universite Paris Diderot, F-91191 Gif-sur-Yvette Cedex, France
         \and 
              Leibniz-Institut für Astrophysik, An der Sternwarte 16, 14482 Potsdam, Germany
         \and
              Institute of Astronomy, University of Cambridge, Madingley Road, Cambridge, CB3 0HA, United Kingdom
         \and
              Unidad de Astronomía, Fac. de Ciencias Básicas, Universidad de Antofagasta,
Avda. U. de Antofagasta 02800, Antofagasta, Chile         
         \and
             Department of Physics and Astronomy, Uppsala University, 751 20 Uppsala, Sweden
         }

   \date{}

 
\abstract 
{We present IFU observations with MUSE@VLT and deep imaging with FORS@VLT of a dwarf galaxy recently formed within the giant collisional HI ring surrounding NGC~5291.
This TDG-like object has the characteristics of typical z=1-2 gas-rich spiral galaxies: a high gas fraction, a rather turbulent clumpy ISM, the absence of an old stellar population, a moderate metallicity and star formation efficiency.}
{The MUSE spectra allow us to determine the physical conditions within the various complex substructures revealed by the deep optical images, and to scrutinize at unprecedented spatial resolution the ionization processes at play in this specific medium.}
{Starburst age, extinction and metallicity maps of the TDG and surrounding regions were determined using the strong emission lines H$\beta$, [OIII], [OI], [NII], H$\alpha$ and [SII] combined with empirical diagnostics. Discrimination between different ionization mechanisms was made using BPT--like diagrams and shock plus photoionization models. }
{Globally, the physical conditions within the star--forming regions are homogeneous, with in particular an uniform half-solar oxygen abundance. At small scales, the derived extinction map shows narrow dust lanes. Regions with atypically strong [OI] emission line immediately surround the TDG. The [OI] / H$\alpha$ ratio cannot be easily accounted for by photoionization by young stars or shock models. At larger distances from the main star--foming clumps, a faint diffuse blue continuum emission is observed, both with the deep FORS images and MUSE data. It does not have a clear counterpart in the UV regime probed by GALEX. A stacked spectrum towards this region does not exhibit any emission line, excluding faint levels of star formation, nor stellar absorption lines that might have revealed the presence of old stars. Several hypotheses are discussed for the origin of these intriguing features. 
}
{}

\titlerunning{ISM properties of a local analogue to high redshift star--forming clumps}

 \keywords{galaxies: dwarf --  galaxies: individual: NGC~5291N -- galaxies: interactions -- galaxies: ISM -- galaxies: starburst -- galaxies: star formation -- HII regions}

 \maketitle 
%

\section{Introduction}

Under the current paradigm of a $\Lambda$CDM cosmology, dwarf galaxies are considered the building blocks of today's massive galaxies \citep{Kauffmann93}, and as such numerous detailed studies are devoted to nearby dwarfs. Among them, the starbursting Blue Compact Dwarf Galaxies (BCDGs), which are less than 1 kpc large and show very low metallicities (1/40th $Z_{\sun}$ < Z < $Z_{\odot}$ / 3), have long been believed to be very young objects \citep{Sargent70, Kunth88}, just like the newly formed galaxies at high redshift. However, the detection of extended red stellar emission \citep{Loose86, Papaderos96} showed that most BCDGs are actually old systems. 

The object investigated in this paper, NGC~5291N (NED\footnote{NASA/IPAC Extragalactic Database.} distance: 63.1 Mpc), was originally classified as a BCDG \citep{Maza91}. It however stands out 
as it shows no hint for an old stellar component \citep{Boquien07,Boquien10c} , is very gaseous -- M$_{gas}$ is around 85 \% of the total mass \citep{Bournaud07} -- and has a clumpy structure.
Its metallicity, about half-solar \citep[][thereafter DM98]{Duc98} is higher than in classical BCDGs, but not so different than that of a number of distant star--forming galaxies with metallicities greater than 0.4 $Z_{\sun}$ at z = 2 \citep{Erb06, Stark08, Steidel14, Zanella15}.
Therefore, the dynamically young dwarf NGC~5291N \citep{Bournaud07} appears as one of the most promising (low--mass) analogue of the distant gas-dominated clumpy galaxies, that are found in number in deep cosmological fields. 
It was formed in a huge gaseous ring (M$_{HI}$ > 10$^{10}$ M$_\sun$, DM98), surrounding the early-type galaxy NGC~5291 that expanded in the intergalactic medium after a violent collision with a bullet galaxy. According to numerical simulations, the encounter occurred 360 millions years ago \citep{Bournaud07}. The HI rich ring formed a series of star--forming clumps. A kinematical analysis \citep{Bournaud07,Lelli15} revealed that three of the gas condensations are most likely gravitationally bound. Our target, NGC~5291N, is the most massive of them. Similar to Tidal Dwarf Galaxies \cite[TDGs, see review by ][]{Duc99}, NGC~5291N has a low dark matter (DM) content. 
As a matter of fact, NGC~5291N shares many properties with the $10^8~\msun$ individual clumps, which formed within the gravitationally unstable gas-dominated disk of distant clumpy galaxies \citep{Zanella15}, and are also presumably dark-matter poor. The stability of such objects against internal feedback is a matter of active debate \citep[e.g. see][]{Genel12, Bournaud14}. It can be investigated in NGC~5291N, with great accuracy, keeping in mind however that the star--formation rate of this local analogue is much lower: around 0.14~$\msun$yr$^{-1}$ for NGSC~5291N \citep{Boquien10c} compared to 32~$\pm$~6~$\msun$~yr$^{-1}$ for the z = 2 clump in \citet{Zanella15}.

The ring of NGC~5291 and its dwarf galaxies within it have been observed by a wide range of ground-based and spatial instruments, resulting to an extensive wavelength coverage: HI 21-cm line with the VLA \citep{Bournaud07}, far--infrared with PACS and SPIRE on Herschel (Boquien et al., in prep.), mid--infrared with Spitzer \citep{Boquien07}, near-infrared with ISAAC on the ESO VLT, optical slit with EMMI on the NTT and H$\alpha$ integral field spectroscopy with a Fabry-Perot (FP) unit on the ESO 3.6m \citep{Bournaud04}, and Far and Near Ultra-Violet with GALEX \citep{Boquien07}. The slit optical spectroscopy of NGC~5291N disclosed a number of bright emission lines, presumably associated with \ion{H}{II} regions; FP observations revealed complex kinematical features within the dwarf on top of the ordered rotation also seen in the HI component. A better understanding of the physical conditions in the interstellar medium requires 3D spectroscopic information, at good spatial and spectral resolution, and a relatively large field of view. This is the capability of the Multi Unit Spectroscopic Explorer (MUSE), which has been recently mounted on the VLT, at Paranal observatory \citep{Bacon10}, and that is about to revolutionize this field of research. We present here MUSE observations of NGC~5291N. They provide us with several ten thousands of resolved spectra and allow us to probe the ionizing processes in the dwarf at physical spatial scales of only 200 pc. Their exploitation is backed by deep multi-band images obtained with the FORS instrument on the ESO VLT.

Former IFU observations of dwarf galaxies have already given us some insight on the ionization properties of these galaxies \citep{Izotov06, Lagos09, Lagos12, Lagos14, James09, James10, James13, James13b} and led to the identification of regions with Wolf-Rayet stars \citep{Cairos10, Kehrig13}, the presence of non-thermal ionization processes such as shocks \citep{Cairos10} or the existence of a central AGN \citep{Cairos09}. 
To our knowledge, this paper is only the second one presenting MUSE observations of a dwarf galaxy, and the first one focused on a colliding system. 
Previously, ionized cones associated with formerly observed Ly$\alpha$ photons leakage, and possibly correlated with outflows, have been identified around the starburst ESO 338-IG04 \citep{Bik15}.

The paper is structured as follows: Sect.~\ref{sec:obs} describes the observation and data reduction procedures. Sect.~\ref{result} presents the detailed analysis of the emission lines and continuum emission, with a focus on their spatial variations. These results are discussed in Sect.~\ref{discussion}. The final conclusions are drawn in Sect.~\ref{conclusion}.

\section{Observation and Data Reduction}
\label{sec:obs}

\subsection{MUSE data}
\label{MUSE}
NGC~5291N was observed using the integral field spectrograph Multi-Unit Spectroscopic Explorer (MUSE) \citep{Bacon10} mounted on the Very Large Telescope (VLT). The observations were performed on June 2014 during the first science verification run (60.A-9320(A), PI: P.-A. Duc). We obtained spectra between 4750~$\AA$ and 9350~$\AA$ over a field of view of 1\arcmin$\times$1\arcmin, which corresponds to 18~kpc$\times$18~kpc at a distance of 63.1~Mpc.

The observations consisted of three on source exposures of 600\,s each, obtained at three position angles (0, 90, and 180$^\circ$). Two exposures of 100\,s were done on an offset (blank) sky field. Twilight sky flats taken the following evening were used in the data reduction process; an exposure of the spectrophotometric standard star LTT\,7987 taken in the following morning twilight was used for flux calibration.
The reduction used the MUSE pipeline \citep{Weilbacher12} through the {\sc EsoRex} program. We used a development version of the pipeline, but the code was very close to the 1.0 release\footnote{Available from ESO via \url{http://www.eso.org/sci/software/pipelines/muse/muse-pipe-recipes.html}.}

We followed the usual steps, with bias subtraction, flat-fielding, and spectral tracing using the lamp-flat exposures, wavelength calibration, all using daytime calibrations made the morning after the observations. The standard geometry table and astrometric solution derived during Commissioning 2a were used to create the intermediate pixel tables. Since the observations were done at a low ambient temperature of $\sim6^\circ$C, slice 6 of channel was not illuminated and could not be traced. We followed the procedure outlined in the pipeline manual to delete the data of this slice from the science pixel tables.

Sky spectra were created using 75\% of the field of view of the two offset sky fields, and the line-spread function derived from the same wavelength calibration exposures was then used to decompose the spectrum into sky emission lines and continuum.

The science reduction was then assigned the sky continuum that was taken closest in time to the science exposure, and we allowed the sky emission line fluxes to be re-adjusted for the science data. The data were corrected for atmospheric refraction, corrected to barycentric velocities (the corrections were about $-23.8$\,km\,s$^{-1}$), before being combined into the final cube. We chose a non-standard sampling in the spectral direction, so that the output cube has steps of $0\farcs2\times0\farcs2\times0.8\,\AA$. The spatial FWHM measured on the white-light image created from this final, combined cube, is about 0\farcs8.\\

\begin{figure*}[]

\includegraphics[width=18cm]{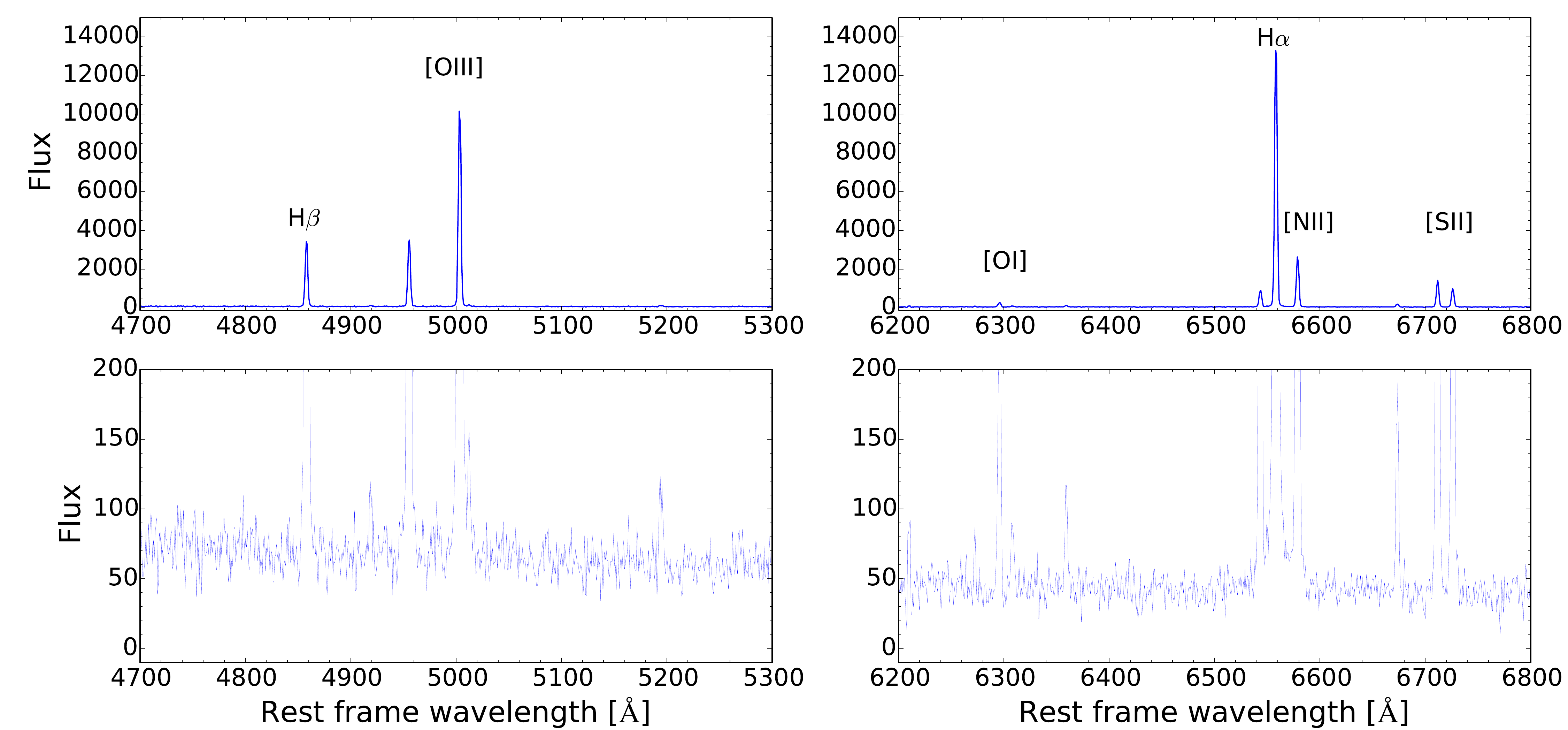}
 \caption{Top: Spectrum of a single spaxel, before the re-binning, at the position of the brightest region with the position of the main emission lines. Bottom: Zoom on the same spectrum. The flux unit is 10$^{-20}$erg.cm$^{-2}$.s$^{-1}$.\label{fig:spectrum}}
\end{figure*}

An example of the extracted spectrum for a single spaxel is shown in Fig.~\ref{fig:spectrum}. It exhibits the very strong emission lines of the Balmer lines H$\beta$ and H$\alpha$, but also forbidden lines such as [OIII] $\lambda\lambda$4959, 5007, [NII] $\lambda$6583, [SII]\footnote{Unless stated otherwise, [SII] refers to [SII]$\lambda6717$ + [SII]$\lambda6731$.} $\lambda\lambda$6717, 6731 and [OI] $\lambda$6300 line. To create reliable spatial maps of line fluxes, the data cube was re-binned by 3x3. Given the depth of the data, using an adaptative smoothing routine was not necessary: for most regions, a signal to noise ratio (S/N) higher than 3 was reached with the adopted fixed 3x3 binning.  The resulting spatial dimensions of the final 3x3 spaxels are $0\farcs6\times0\farcs6$, or 180~pc$\times$180~pc.

The continuum emission was extracted from the spectra with the \textsc{pyraf} {\it continuum} procedure. 
Emission line fluxes were computed with custom-made Python scripts based on Gaussian fits. We only studied regions where the emission was 3$\sigma$ above the noise and we made sure that our Gaussian fitting error was less than 10\% of the total line flux. Unfortunately, the redshift of the galaxy put the [SII]$_{\lambda6731}$ very close to airglow lines at 6827~$\AA$, making it very hard to detect at low luminosity. However, the [SHII]$\lambda$6717/[SII]$\lambda$6731ratio is a tracer of the electron density. In the low-density limit, it reaches 1.45. In the brightest region in our field, the lowest value is 1.15, corresponding to an electron density of 300~cm$^{-3}$. Therefore, in the case of a non-detection of [SII]$_{\lambda6731}$ but a good detection of every other lines, a value of [SII]$_{\lambda6731}$ = [SII]$_{\lambda6717}$/1.42 is chosen, which corresponds to a maximum error of 9$\%$ on the value of [SII] = [SII]$_{\lambda6717}$ + [SII]$_{\lambda6731}$ for the densest regions, which is lower than the uncertainty on the Gaussian fit, and less than 2$\%$ for regions with $n_\mathrm{e}$ < 100~cm$^{-3}$ which are the hardest to resolve. The uncertainty for the value of [SII]$_{\lambda6731}$ is not available for the spaxels which needs this method, as it is dominated by the noise from the sky line. When mean uncertainties involving the [SII] emission doublet are to be assessed,  points where [SII]$_{\lambda6731}$ is blended by the airglow line will not be taken into account.

\subsection{FORS data}
The deep optical V, R and I images have been acquired in March 2010 with the instrument FORS on the VLT (Program 382.B-0213(A), PI: E. Zackrisson). The final images, covering a total field of view of 11.2\arcmin $\times$ 15.5\arcmin were obtained by registering and combining 26 individual exposures of 300 sec, 300 sec and 240 sec in the V, R and I Bessel filters respectively. Photometric zero-points were provided by the ESO pipeline. The limiting surface brightness in the V band is estimated to 27 mag arcsec$^{-2}$ (Vega), from the local fluctuations in the background. 
The average FWHM of point sources in the stacked image is 1.0\arcsec.\\

\begin{figure*}[]
\includegraphics[width=19cm]{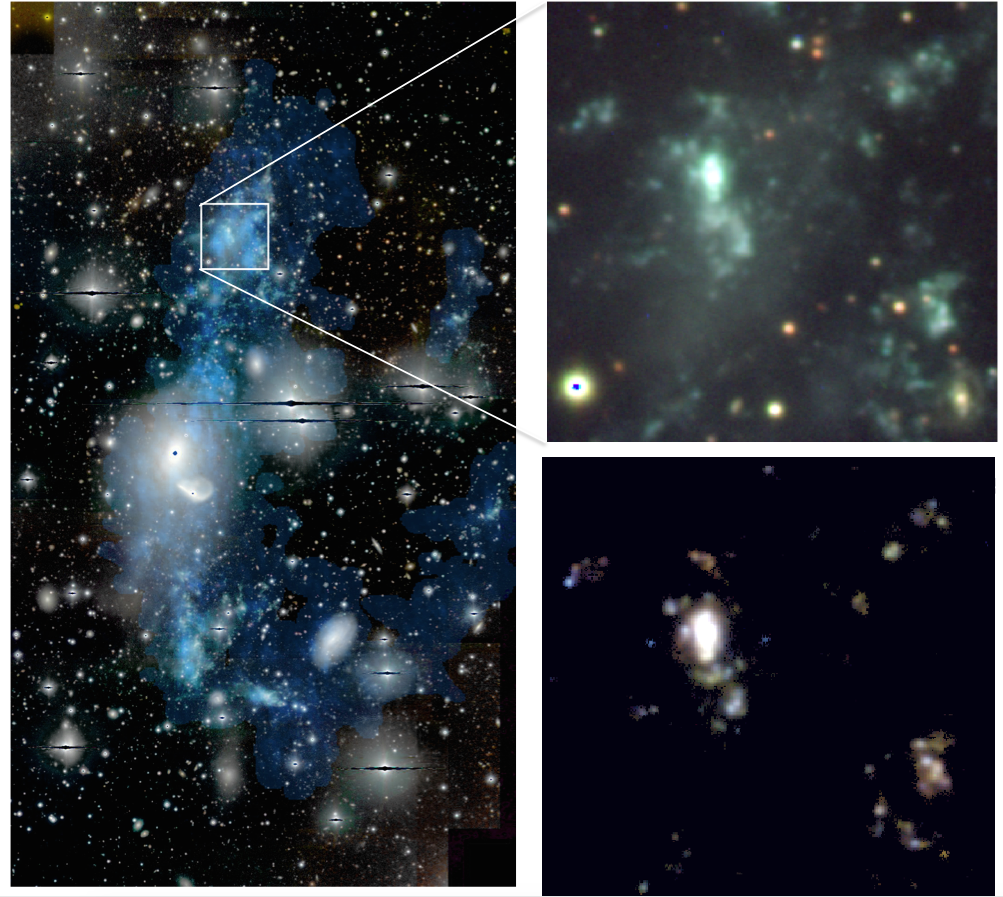}
 \caption{Left: Composite image of the NGC~5291 collisional ring. Background is a FORS composite color image obtained with the V, R, and I bands. HI data from the VLA is superimposed in blue. North is up and East is left. The white square centred on NGC~5291N delineates the MUSE field of view. Upper right: FORS V, R, I composite color image of the MUSE target. Lower right: MUSE composite color image using three different emission lines: {\it Red:}[NII]$\lambda$6583, {\it Green:} H$\alpha$, {\it Blue:} [OIII]$\lambda$5007. Each pixel covers a 60~pc~$\times$~60~pc region and the full image corresponds to 1\arcmin$\times$1\arcmin, or about 18 kpc x 18 kpc at the distance of NGC~5291N, 63.1~Mpc.\label{Picture}}
\end{figure*}

The FORS image of the whole system, including the collisional ring and its host galaxy, is shown on the left part of Fig.~\ref{Picture}, with the HI emission from the VLA superimposed in blue. The radio data highlights the shape of the ring structure. The field of view of the MUSE observations towards the TDG NGC~5291N is delineated by the white square.
Both the composite FORS broad-band image (Fig~\ref{Picture}, top-right) and the MUSE narrow-band image (Fig.~\ref{Picture}, bottom-right) reveal the clumpy structure of the ISM of NGC~5291N. 
The FORS image mostly exhibits continuum emission; the MUSE image is a combination of emission from three main lines: [OIII] ($\lambda$5007), H$\alpha$. and [NII] ($\lambda$6583). As shown by comparing the FORS and MUSE images (see also Fig.~\ref{Picture}), the individual clumps are associated to strong emission lines and are embedded in regions with extended diffuse continuum emission.

\section{Results\label{result}}
As illustrated in Fig.~\ref{Picture}, the MUSE spectrophotometric data show a striking large heterogeneity towards a recently born object, NGC~5291N, which could a priori be considered simple. While the continuum emission has a rather uniform blue color, the emission line ratios strongly vary at small scales. We analyse in the following the spatial variations of the observed spectral lines and investigate various diagnostics to account for them. 

\subsection{Dust extinction}

Before investigating the spatial variations of the emission line ratios, dust extinction maps were computed to allow for reddening correction of the emission line fluxes. These maps were derived from the H$\alpha$/H$\beta$ Balmer line ratio. The LMC extinction law from \citet{Gordon03}, with $R_{V}$ = 3.41, was used in combination with a theoretical value for the unobscured line ratio for case B recombination of H$\alpha$/H$\beta$ = 2.86, for $T_e$ = 10,000 K and $n_e$ = 100~cm$^{-3}$ \citep{Osterbrock89}. 

\begin{figure}[]
\includegraphics[width=9.6cm]{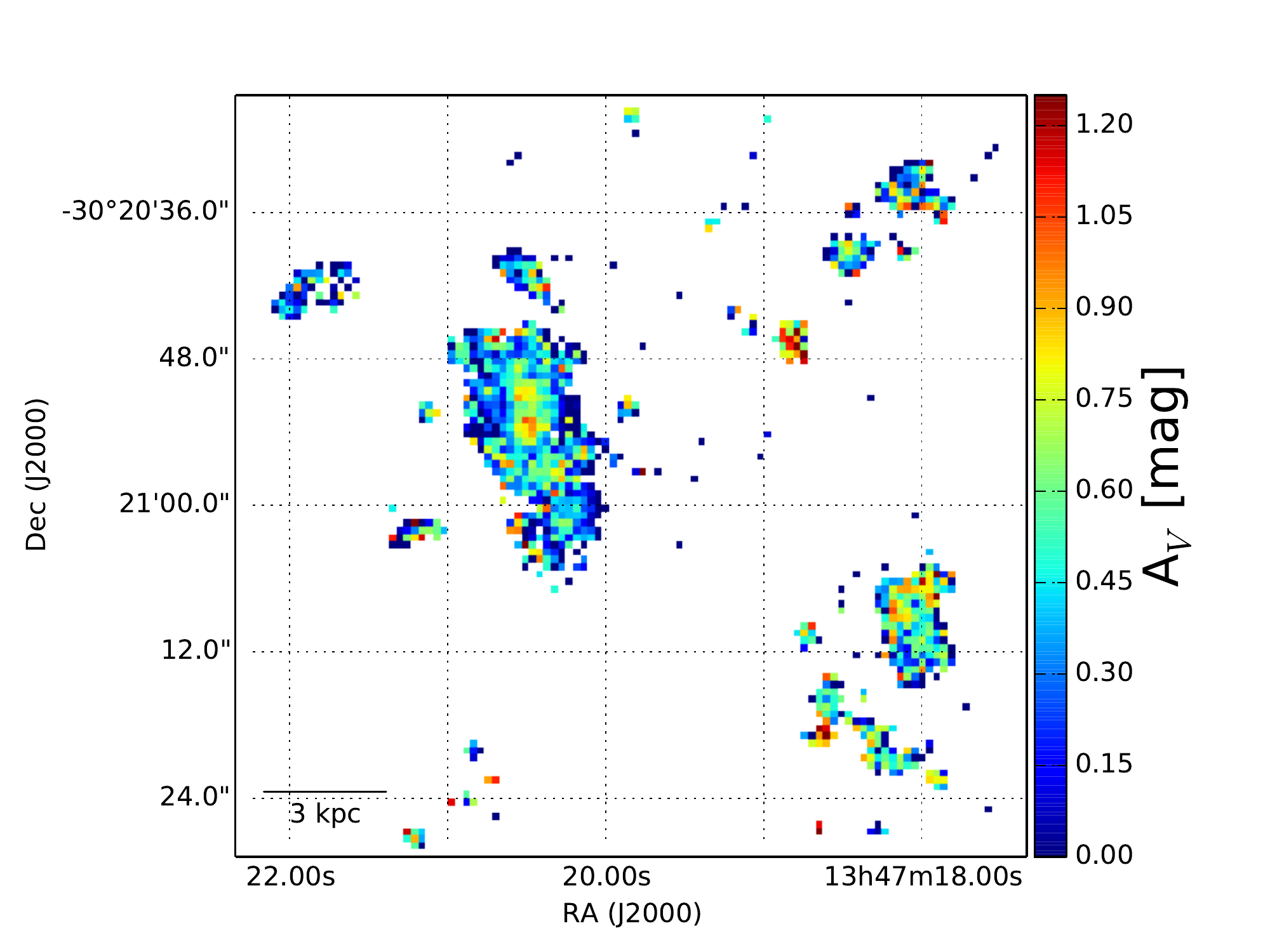}
\caption{Spatial distribution of the dust extinction, A$_V$. The dust is heterogeneously distributed in the system, but its distribution peaks where star formation is the most recent, see Fig.~\ref{Hb_EW}.\label{fig:att}}
\end{figure}

\begin{figure}[]
\includegraphics[width = 9.4cm]{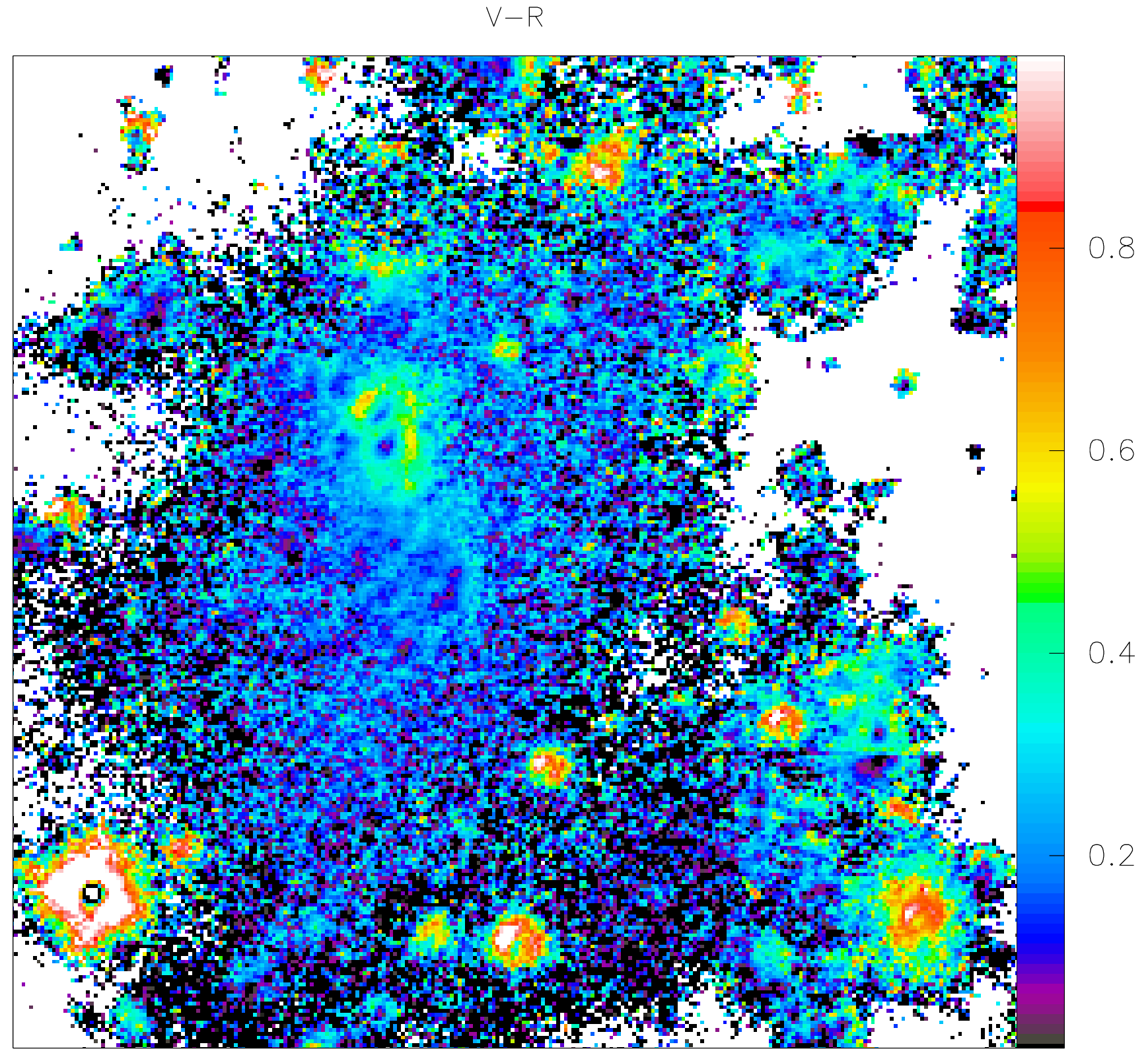}  
\caption{V-R color map of NGC~5291N derived from the FORS deep image, scaled in mag. The field of view is the same as in Fig.~\ref{fig:att}.\label{fig:V-R}}
\end{figure}

As shown in Fig.~\ref{fig:att}, the extinction is globally low with however local variations of Av by up to 1 mag. The dust reddening peaks at a value of $A_V$ $\simeq$ 1.25~mag towards regions where dust emission had been observed with Spitzer and Herschel. The FORS color image also show the presence of narrow dust lanes at this location (Fig.~\ref{fig:V-R}). The extinction derived from the MUSE data is consistent with the one derived by \cite{Boquien10c} from their fit of the Spectral Energy Distribution probed by the FUV, NUV, B, V, R, J,  H, and K bands. It is higher than that originally estimated by DM98 from slit spectroscopy.
 
In the following all emission line fluxes are corrected for dust extinction.

\subsection{Spatial Variation of the Emission Line Ratios}

\begin{figure*}[]
\includegraphics[width=18cm]{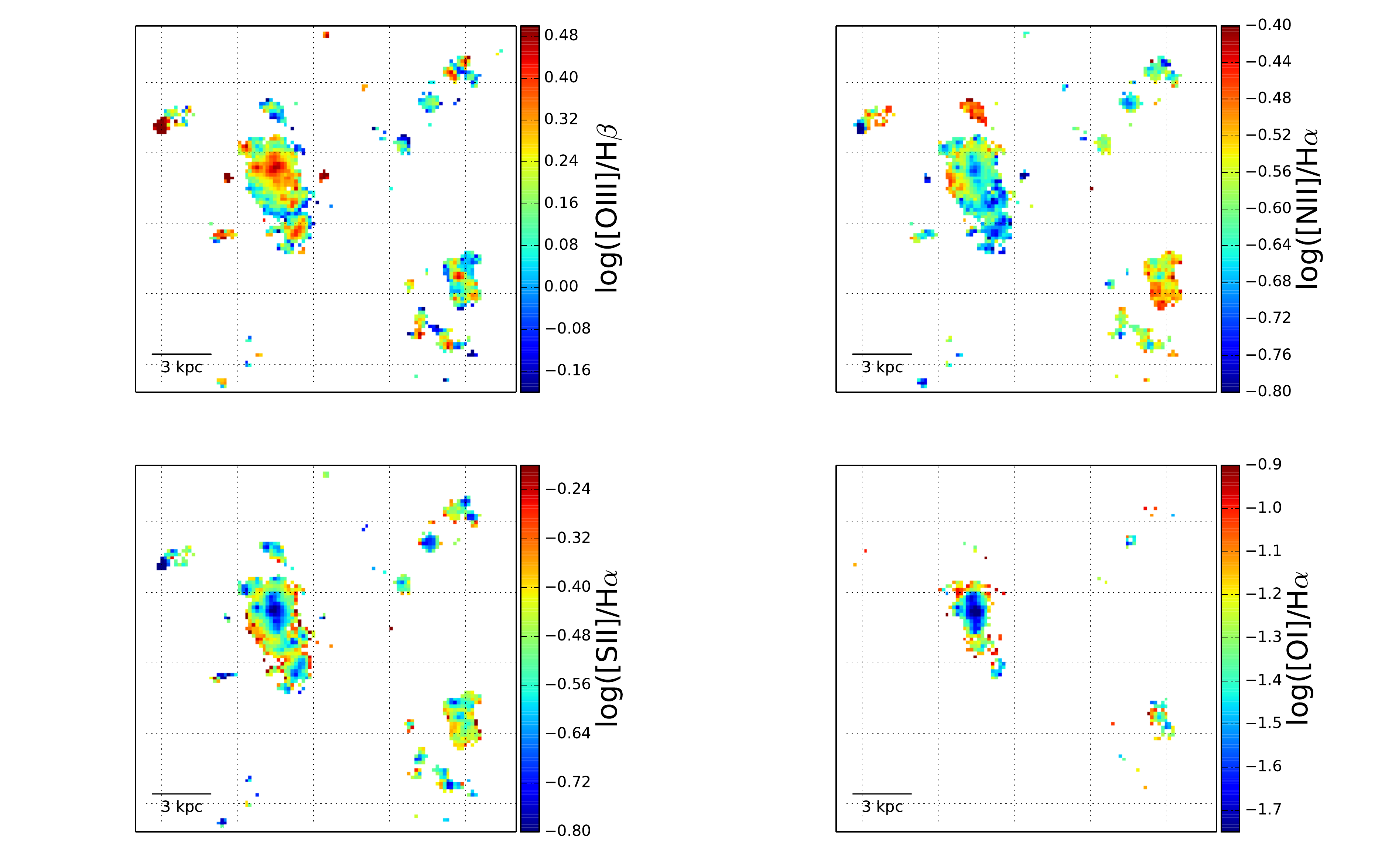}
\caption{Spatial distribution of the principle emission-line flux ratios. A logarithmic scale is used for coding the line ratios. The field of view is the same as in Fig.~\ref{fig:att}.
 \label{ratio_map}}
\end{figure*}

The spatial variations of the main emission line ratios are plotted in Fig.~\ref{ratio_map}. An inside-out positive gradient is seen in the [NII]$\lambda$6584 / H$\alpha$, [SII]$\lambda\lambda$6717,6731 / H$\alpha$ and [OI]$\lambda$6300 / H$\alpha$ maps whereas the gradient is negative for the [OIII]$\lambda$5007/H$\beta$ line ratio. 
Besides these large scale variations, some clumps exhibit deviant emission line ratios (see also the lower right panel of Fig.~\ref{Picture}).
 The one immediately North of the TDG has the strongest [NII]/H$\alpha$ ratio in the field and the clump to the North-West the strongest [OIII]/H$\beta$ ratio. 
 
\begin{figure*}[]
\includegraphics[width=18cm]{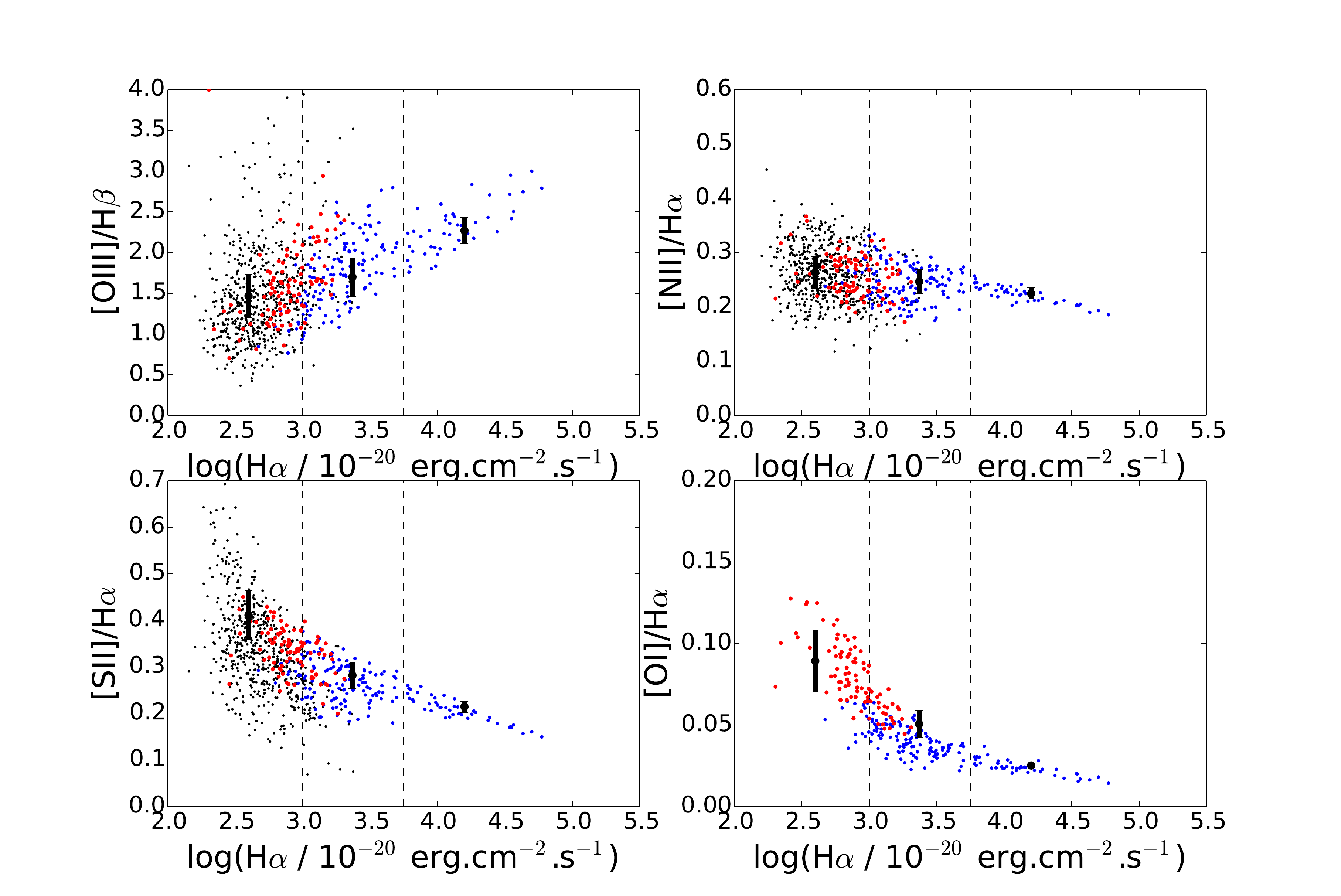}
\caption{Distribution of emission-line ratios as a function of the H$\alpha$ flux, also tracing the local H$\alpha$ flux surface brightness. The blue resp. red points correspond to those located inside resp. outside the starburst locus in the [OI]/H$\alpha$ BPT diagnostic diagram (see Fig.~\ref{fig:BPT}). The small black points correspond to spaxels with undetected [OI] line. The error bars are 1$\sigma$ typical dispersion and their y-axis position is the mean value of the ratio in the given H$\alpha$ flux range.\label{fig:dig_cumul}}
\end{figure*}

 The flux ratios also depend on the flux intensity of the lines.
 In Fig.~\ref{fig:dig_cumul}, the emission line ratios are plotted as a function of the H$\alpha$ flux. [SII]/H$\alpha$ and [OI]/H$\alpha$ ratios increase significantly with lower H$\alpha$ surface brightness, whereas [OIII] / H$\beta$ decreases.
 
 The scatter is large at low H$\alpha$ surface brightness, partly due to the increase of the measurement errors. 

The ISM can be ionized by several mechanisms: photoionization due to ultraviolet (UV) photons from young, hot stars, photoionization from an Active Galactic Nucleus (AGN), shock ionization due to stellar winds and supernovae, or shock ionization from dynamical processes such as collisions. In order to gain insight into the processes at play, line ratio diagnostics diagrams, known as BPT diagrams \citep{Baldwin81, Veilleux87} are used: [OIII]$\lambda$5007 / H$\beta$ versus [NII]$\lambda$6584 / H$\alpha$, [SII]$\lambda\lambda$6717,6731 / H$\alpha$ and [OI]$\lambda$6300 / H$\alpha$. Although BPT-like diagrams were originally applied to  central galactic regions or even to entire galaxies to diagnose AGN or LINER emission, they are used here at the scale of a spaxel to identify the different ionizing processes. 

\begin{figure*}
\includegraphics[width=18cm]{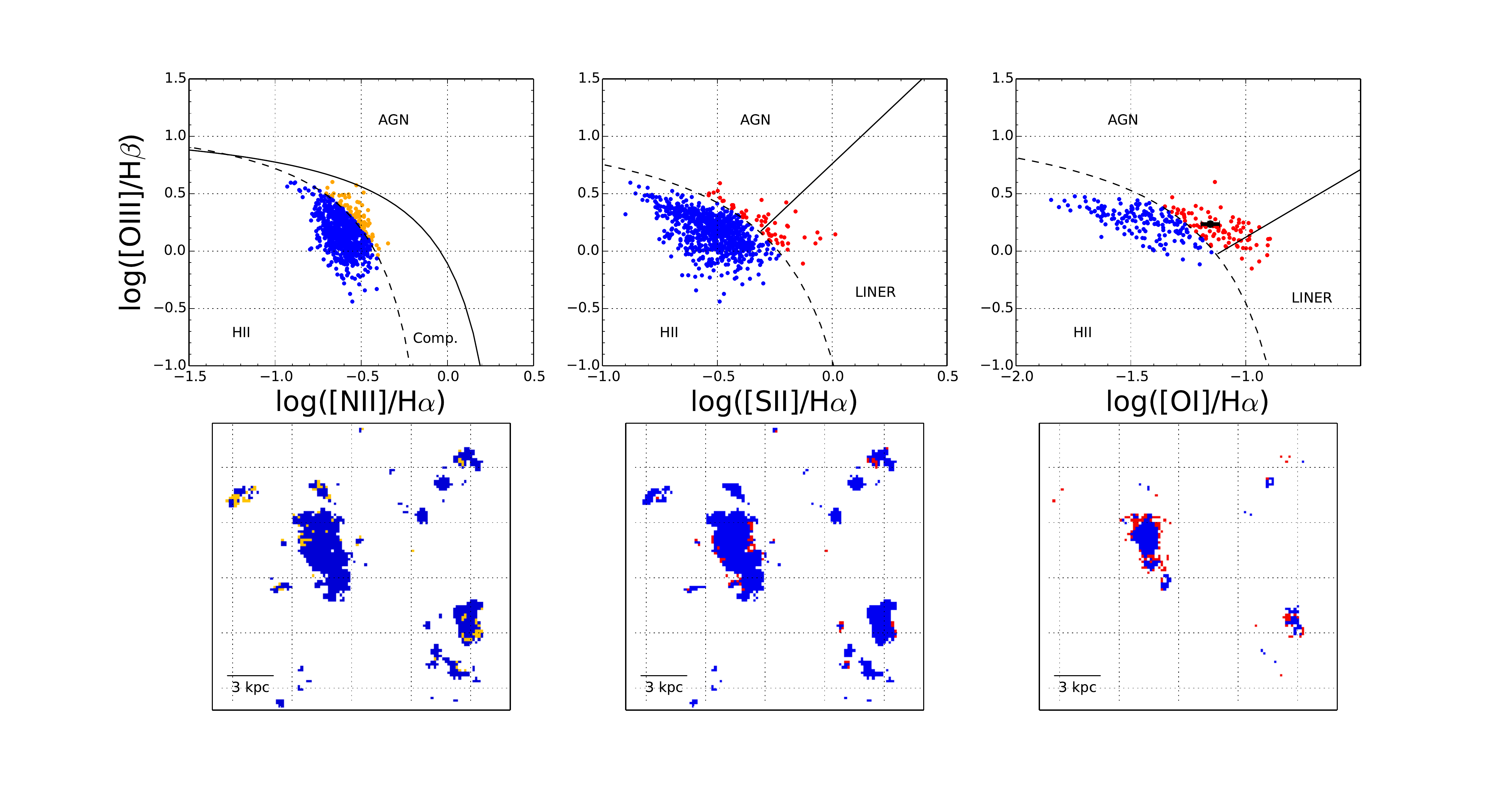}
 \caption{BPT diagnostics for different emission line ratios. On the [OIII]/H$\beta$ vs [NII]/H$\alpha$ diagram, the dotted line delineates the pure starburst region, as defined by \citet{Kauffmann03} using a set of Sloan Digital Sky Survey (SDSS) spectra.
The spaxels belonging to this region are shown in blue, those outside it in orange; their spatial location is shown in the figure below. The field of view is the same as in Fig.~\ref{fig:att}.
The solid line traces the upper theoretical limit to pure \ion{H}{II} regions measured by \citet{Kewley01}.
For the other diagrams, the dashed lines defined by \citet{Kewley06} separate the starburst (blue points) from the AGN/Liner (red points) regions. The solid lines further distinguish between AGN ionization (above the line) and LINER ionization (below).
The corresponding spatial distribution of starburst/non-starburst 3$\times$3 binned spaxels are shown below. 
The highest fraction of spaxels inconsistent with a photoionization by a starburst show up on the  [OIII]/H$\beta$ vs [OI]/H$\alpha$ diagram. Their spectra have been stacked and the resulting extracted line ratios are shown with the black point, together with the error bar.
\label{fig:BPT}} 
\end{figure*}

Emission line ratios for individual rebinned spaxels are plotted on the BPT diagnostic diagrams shown in Fig.~\ref{fig:BPT} .
 Most of our data points fall in the locus of star forming regions for two diagnostic diagrams. However, the distribution on the [OI]/H$\alpha$ diagnostic does not look standard. It shows a high number of points located outside of the locus for classical \ion{H}{II} regions. As shown in Fig.~\ref{fig:BPT}, bottom, the deviant points are located in the outskirts of the star--forming regions. The mean uncertainty on the value of log([OI] /H$\alpha$) for the red points is 0.19, which is not enough to explain their location in the AGN part of the BPT diagram. One should also note that the stacked spectrum of these regions is clearly located in the AGN region of the diagram, with very little uncertainty on the measure.

We investigate in the following the possible causes of the spatial variations of the emission line ratios, starting from the innermost regions to the external ones.

\subsection{Star--forming regions}

 \begin{figure}
\centering\noindent

\includegraphics[width=9.6cm]{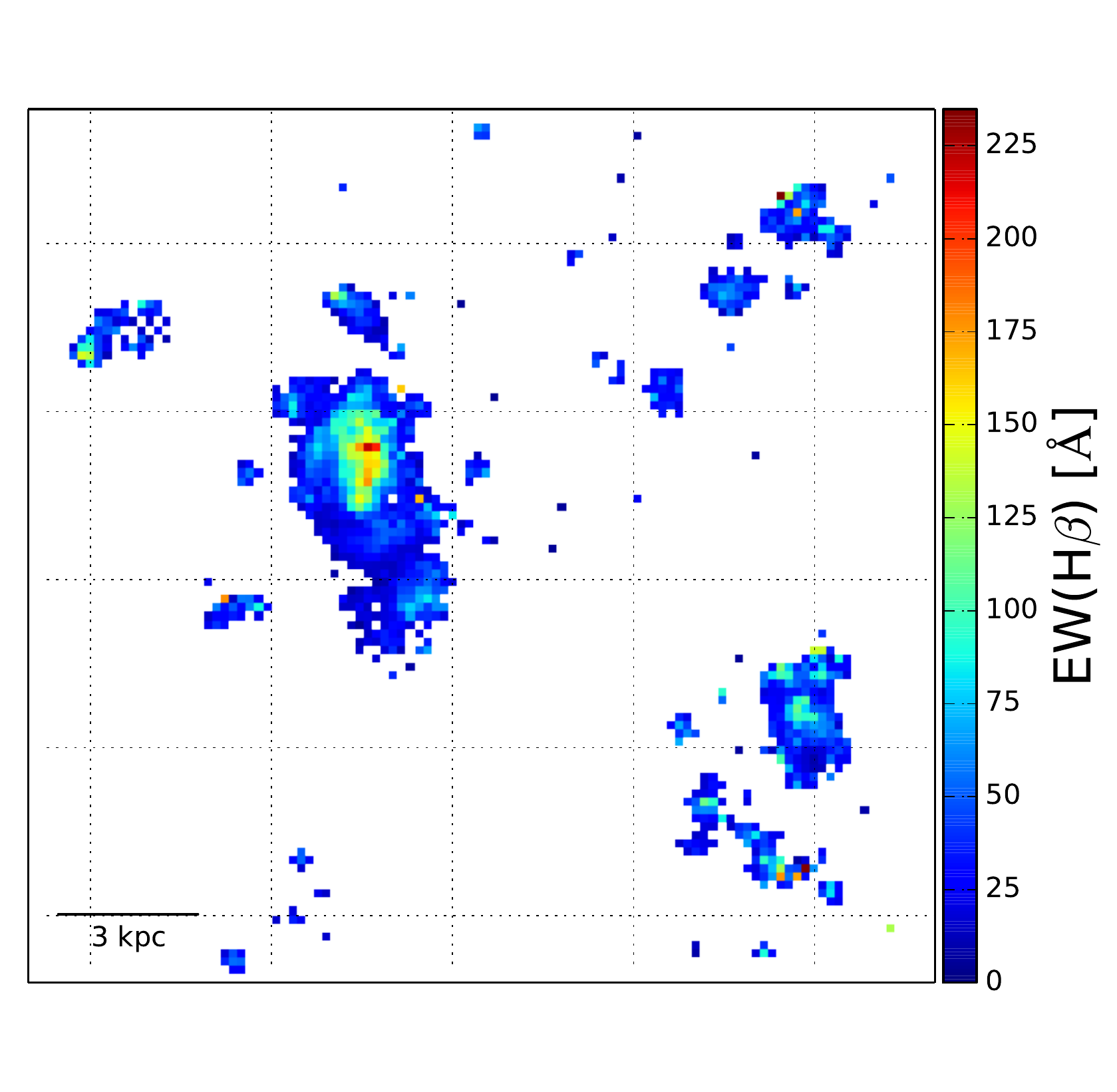}
 \caption{Spatial distribution of the equivalent width of H$\beta$. The field of view is the same as in Fig.~\ref{fig:att}.\label{Hb_EW}}

 \end{figure}

The equivalent width of H$\beta$ (EW(H$\beta$)) gives constraints on the age of a starburst episode \citep{Stasinska96, Zackrisson01, Terlevich04}. 
The spatial distribution shown in Fig.~\ref{Hb_EW} peaks at 230~$\AA$, in the centre of the brightest region. EW(H$\beta$) decreases further out to 30~$\AA$ in the outskirts of the dwarf.
Following \citet {Boquien07}, PEGASE II models \citep{Fioc97, Fioc99} were used to estimate the ages of the star--formation episodes. Assuming an instantaneous starburst and a constant metallicity Z = 0.008 (see Sect.~\ref{met}), we obtained starburst ages of 3 Myr for the central region and 7 Myr for the oldest regions. Under the assumption that the underlying continuum is only due to the newly formed stars (see Sect.~\ref{sect:emission} and ~\ref{sect:emission_discussion}), the EW(H$\beta$) map looks consistent with an outside-in star--formation episode. 

Given the very young age of the central starburst, the presence of Wolf-Rayet (WR) stars is expected. Indeed, DM98 detected a WR bump in their long-slit spectra of NGC~5291N. However, our spectral window does not allow us to reach the location of the bump: the blue boundary of our spectral range is 4682~$\AA$ in rest-frame referential whereas the WR bump peaks at 4686~$\AA$. A rising at the far blue end of the spectrum was glimpsed, consistent with the WR feature, but could not be discriminated from an observational bias. Furthermore the red WR bump around 5650 to 5800~$\AA$ is not observed in the MUSE data.

\subsection{Metallicity}
\label{met}

The chemo-dynamical simulations of \cite{Ploeckinger14} predict some level of self-enrichment in Tidal Dwarf Galaxies which should translate into spatial variations of the metallicity and thus contributes to the scatter observed in each BPT diagram of Fig.~\ref{fig:BPT}.

We derived the metallicity from the strong emission lines ratios. Unfortunately, our spectral range does not include the [OIII]${\lambda4363}$, so we could not compute the electron temperature, needed for a direct measurement of the metallicity. 

 Instead, we estimated the metallicity from the empirical calibrations based on the parameter O3N2 = $\log_{10}\left(\frac{[OIII]\lambda5007 / H\beta} {[NII]\lambda6548 / H\alpha}\right)$.
 Initial estimates by DM98 indicated rather high values -- about half solar -- for the oxygen abundance. The \cite{Marino13} empirical calibration was then chosen. Indeed, compared to other calibrations using this parameter \citep[e.g.][]{Pettini04}, it includes high metallicity \ion{H}{II} regions and should provide better estimates for high Z regions. Following their calibration:
\begin{equation}
12 + \log_{10}(O / H) = 8.55 - 0.221 \times O3N2
\end{equation}
 for O3N2 < 1.7, we get <12 + log(O/H)> = 8.38 with a standard deviation $\sigma$ = 0.05, which corresponds to an abundance of Z~$\simeq$~0.5~Z$_{\odot}$, using 12 + log(O/H)$_{\odot}$ = 8.69 \citep{Asplund09}. This calibration has a typical uncertainty of $\pm$ 0.2 dex.

\begin{figure}[]
\includegraphics[width=9cm]{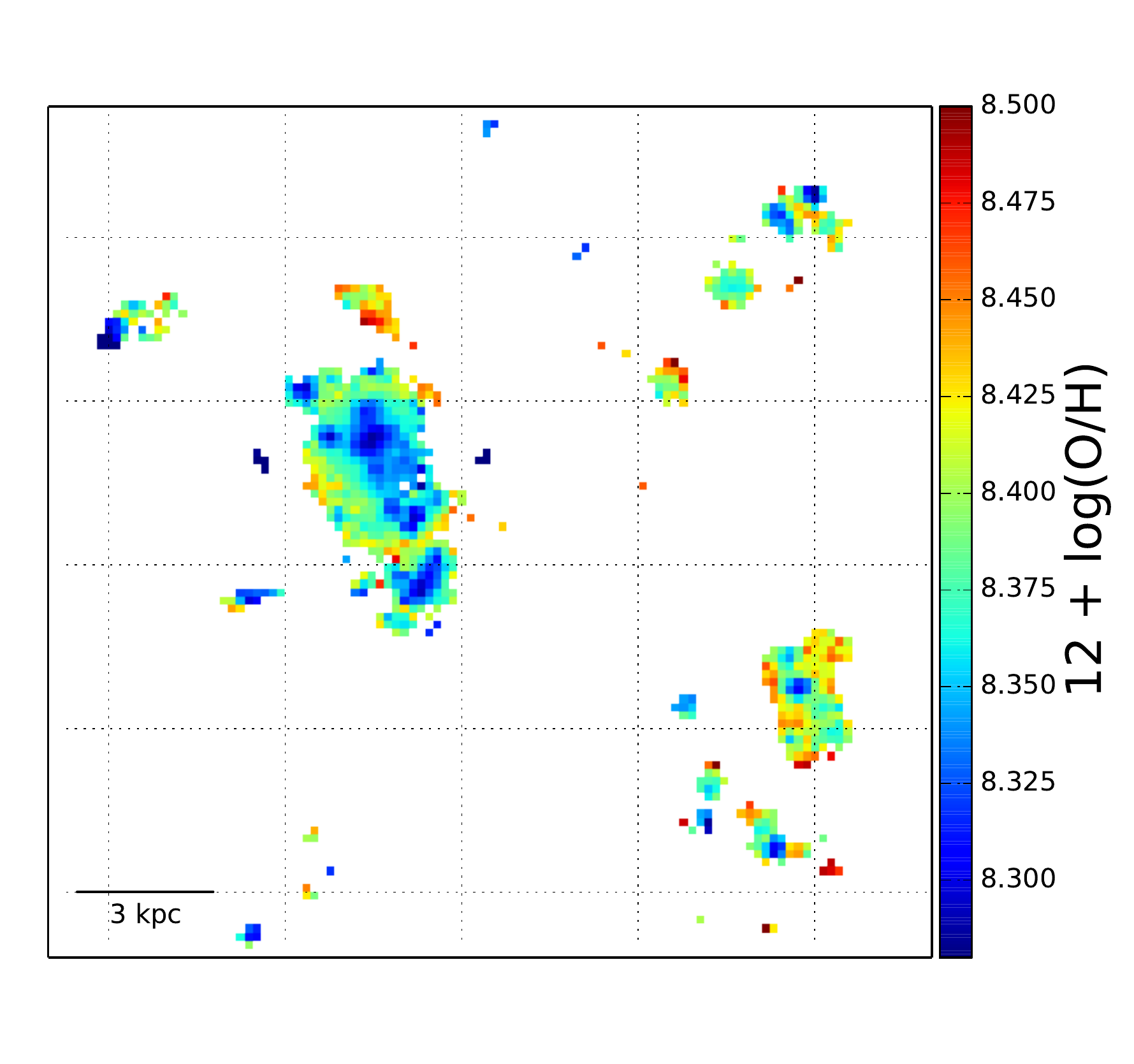}
 \caption{Metallicity map estimated from O3N2 and the calibration of \citet{Marino13}. An outside-in gradient is observed, but the difference between the lowest and the highest abundance can be consistent with an homogeneous abundance distribution and be caused by the variation of other parameters, such as the ionization parameter.The field of view is the same as in Fig.~\ref{fig:att}.\label{metallicity}}
\end{figure}

The spatial distribution of the metallicity is shown in Fig.~\ref{metallicity}. An apparent decrease of the metallicity seems to be observed towards the brightest and dustier regions. This is at odds with the simulations of \citet{Ploeckinger14} that predict a higher metallicity where star--formation is the most active. 

However, our positive metallicity gradient is of 0.2 dex, i.e. within the uncertainty of the method we used. So it may in fact corresponds to variations of other physical parameters such as the ionization parameter, and the data are consistent with a homogenous metal distribution within NGC~5291N. 
One should note that previous studies, such as DM98, detected no metallicity gradient over the whole HI ring surrounding NGC~5291, whose radius is around 200 kpc and which contains several other star--forming condensations. 

\subsection{Ionization Parameter}
\label{coeff_ion}

\begin{figure*}
\includegraphics[width=18.5cm]{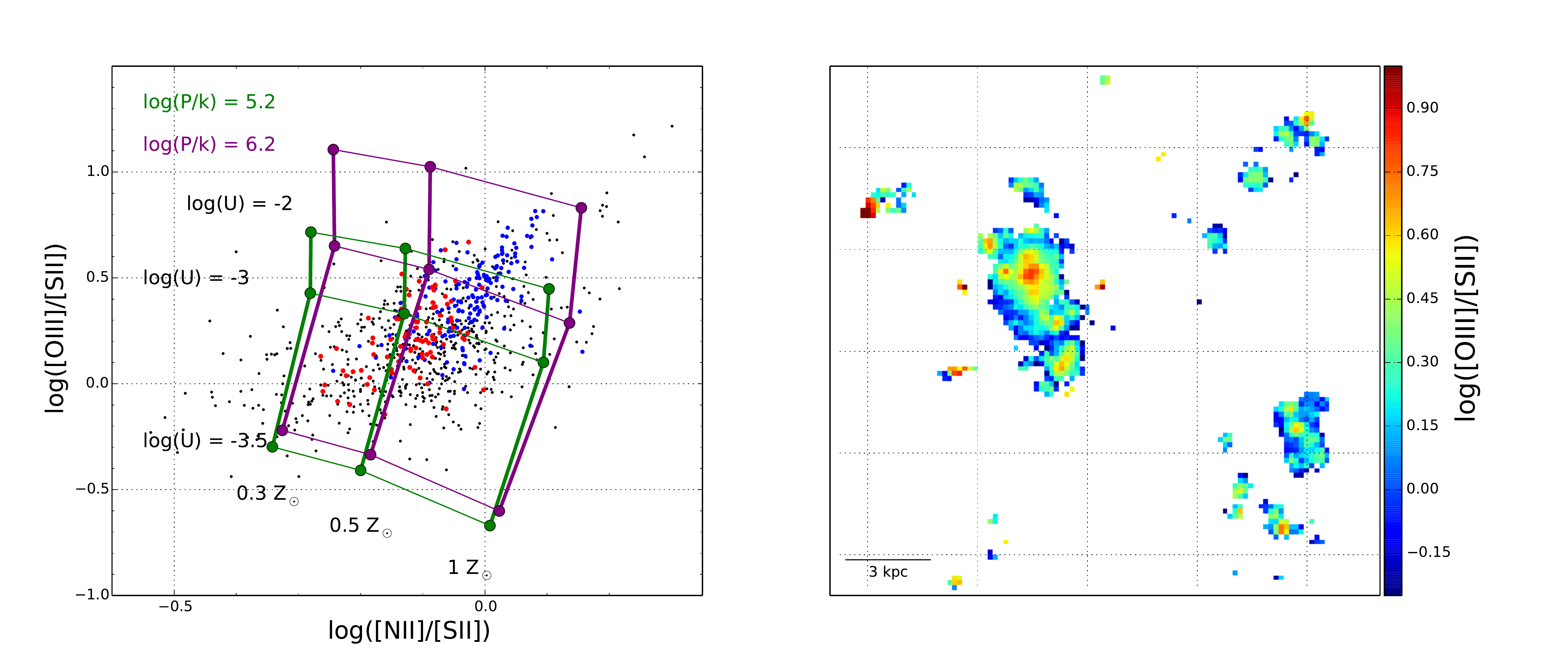}

 \caption{Left: Metallicity-ionization parameter grid from the MAPPINGS V ionization model. The red points are the ones falling outside of the SF region in the third BPT diagram. The colored dashed lines represent the grid for a constant metallicity and the continuous black lines for a constant ionizing parameter. The dimensionless ionizing parameter goes from log(U) = -3.5 to log(U) = -2. Right: Spatial distribution of the logarithmic values of the [OIII]/[SII] ratio, used as a proxy for the ionization parameter. We see an outside-in gradient showing that the most ionized regions are located in the centre of the galaxy. The field of view is the same as in Fig.~\ref{fig:att}. \label{fig:axion}}
\end{figure*}

The ionization parameter q\footnote{The dimensionless ionizing parameter U = q/c, where c is the speed of light, is used in the following.}, defined as the ratio between the flux of ionizing photons and the number density of hydrogen atoms, also has an influence on the values of the emission line ratios \citep{Dopita13}. To determine the distribution of the ionizing parameter throughout the system, the grids from the photoionization model MAPPINGS V\footnote{Available at \url{miocene.anu.edu.au/Mappings}} \citep{Sutherland15} are used on a [OIII]/[SII] versus [NII]/[SII] diagram, which offers a convenient way to break the degeneracy between the metallicity and ionizing parameter. The electronic density, computed from the [SII]$_{\lambda6717}$ / [SII]$_{\lambda6731}$ ratio, reaches a maximum value of 300~cm$^{-3}$ (see Sect.~\ref{MUSE}) and can get as low as 9cm$^{-3}$ (see Sect.~\ref{Shocks}). Moreover, the electronic temperature was estimated to be 12700~K by DM98. Two grids were then used, with respectively log(P/k) = 5.2 and 6.2 to cover the density range encountered in the galaxy. The shape of the ionizing Extreme Ultra-Violet photon spectrum comes from the STARBURST99 synthesis model \citep{Leitherer99}, assuming a Saltpeter IMF, \citet{Lejeune97} atmospheres and a continuous star--formation extending over 4 Myr, as described in \citet{Dopita13}. The left panel of Fig.~\ref{fig:axion} shows that the ionizing parameter values are quite dispersed, going from log(U) = -3.5 to log(U) = -2. The spatial distribution of the [OIII] / [SII] emission line ratio, which can be used as a proxy for the ionization parameter, is plotted on the right panel of Fig.~\ref{fig:axion} and shows similar gradient as the one observed in Fig.~\ref{ratio_map} and can therefore be accounted for the spread observed on the three BPT diagrams.

\subsection{Continuum emission\label{sect:emission}}

\begin{figure}[]
\includegraphics[width=8.6cm]{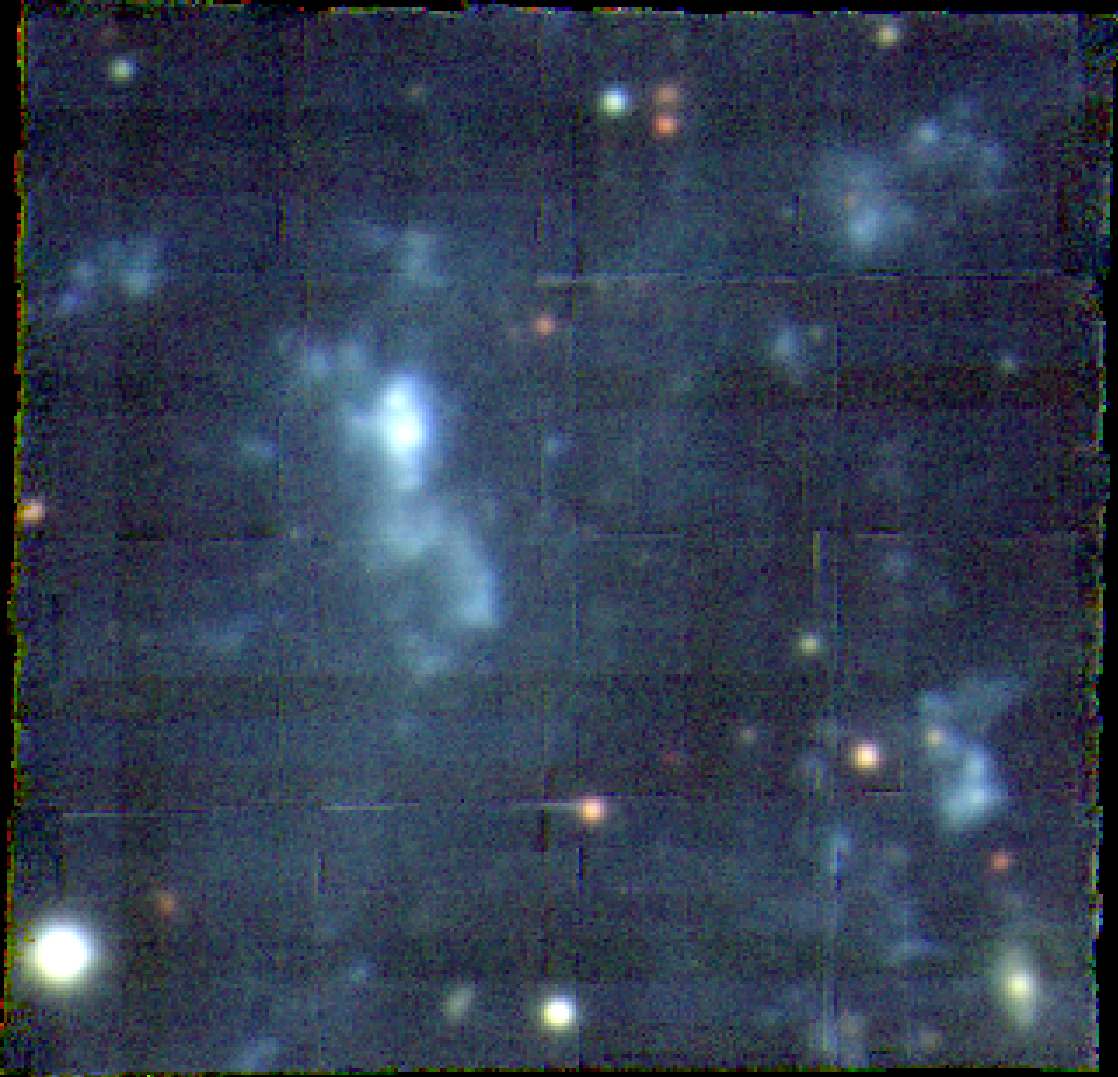}
\caption{Reconstructed composite color image of the extracted continuum of NGC~5291N using similar bands as V, R and I on MUSE. The blue diffuse continuum emission is detected, although less clearly than on the FORS image presented on the upper right panel of Fig.~\ref{Picture}. The image was created from the un-binned data-cube.\label{cont_RGB}} 
\end{figure}

\begin{figure}[]
\includegraphics[width=9cm]{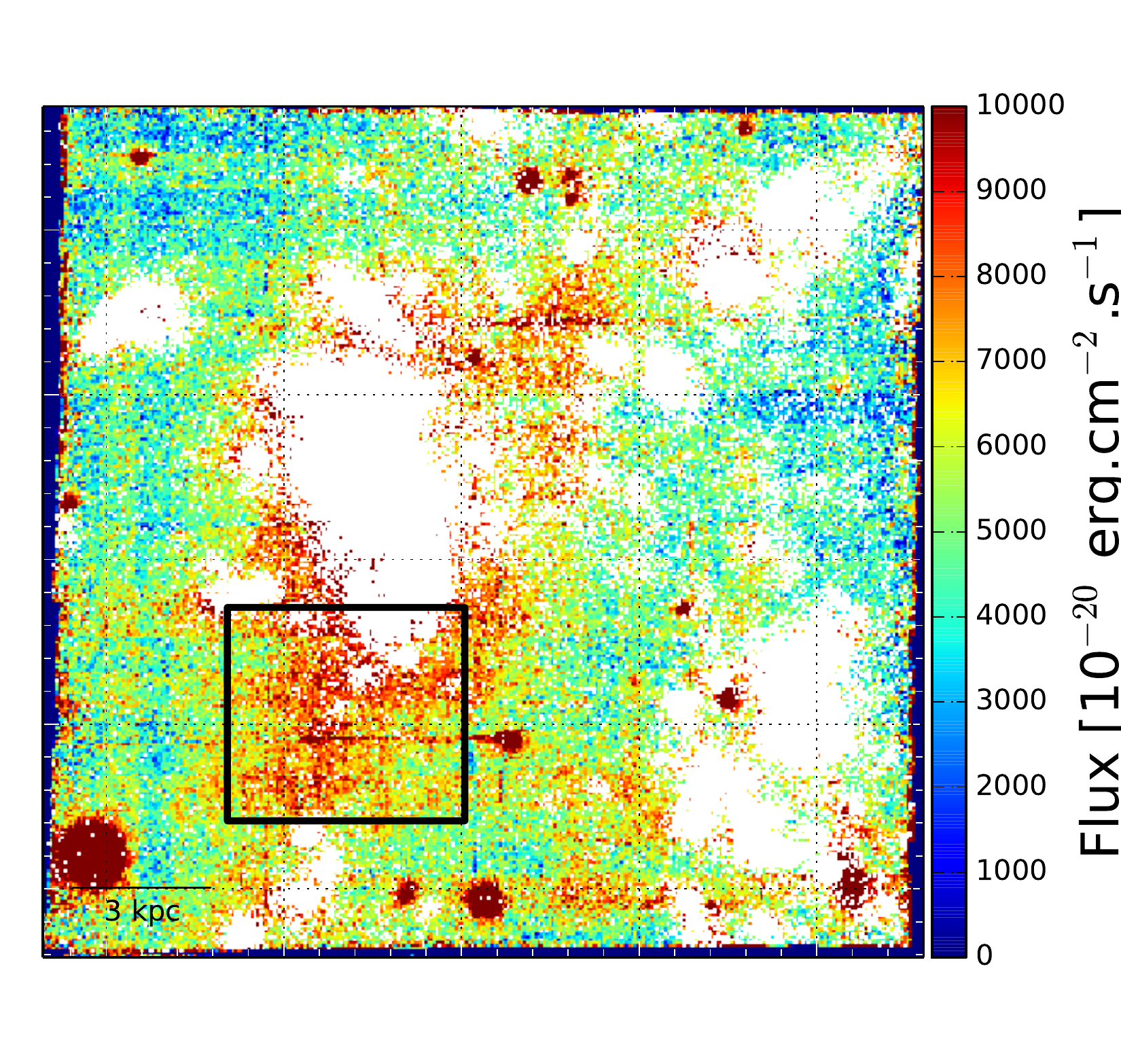}
\caption{Integrated emission of the continuum in 10$^{-20}$~erg~s$^{-1}$~cm$^2$ between 4750~$\AA$ and 6000~$\AA$. Spaxels for which a 3$\sigma$ H$\alpha$ emission was detected are shown in white. The black rectangle indicates the region whose spectra were combined to obtain the stacked spectra of Fig.~\ref{cont_spectra}. The field of view is the same as in Fig.~\ref{fig:att}. \label{fig:cont}} 
\end{figure}

Besides the emission lines, the continuum emission from the MUSE spectra was also extracted, and its spatial distribution was investigated. Away from the main star--forming regions, this continuum emission is quite faint and hardly above the noise level. As shown in Fig.~\ref{fig:V-R} and Fig.~\ref{cont_RGB}, this diffuse emission is spatially extended and rather blue (V-R $\simeq$ 0.2~mag). 
 The existence of this diffuse component is secured by the deep broad band images obtained with FORS: see the similarity between the reconstructed MUSE map of Fig.~\ref{cont_RGB} and the composite image of FORS shown on the upper right panel of Fig.~\ref{Picture}. 
Fig.~\ref{fig:cont} further shows that this component extends in areas where no strong emission lines, in particular H$\alpha$, is observed. 

\section{Discussion}
\label{discussion}

The target investigated here, NGC~5291N, is supposedly a ``simple`` object. It was formed less than a few hundred Myr ago, following a high speed collision, in an initially purely gaseous structure \citep{Bournaud07}. Yet the FORS deep images and MUSE data presented here already show a complex interstellar medium. The ISM in this still gas-dominated object appears to be very clumpy, like the typical star--forming galaxies at $z>1$.
The high spatial resolution and large field-of-view of MUSE allowed us to map the flux distribution of multiple emission emission lines, as well as the flux ratios between them up to large radial distances.  

 Large and small scale variations of emission line ratios are observed. In particular the [OI]/H$\alpha$ ratio has excursions beyond the locus of typical starbursts. Deviant points surrounds the main star--forming regions and are thus found at relatively low H$\alpha$ surface brightness.
Such variations of the emission lines cannot be due to local changes of the dust extinction or metallicity, which we found to be pretty uniform. We discuss here other possible origins for the deviant line ratios.

\subsection{Origin of the strong [OI]/H$\alpha$ emission}

The [OI]$\lambda$6300 is a faint line, making if difficult to map. However, the sensitivity of MUSE, allowed us to detect it with a S/N $>$ 3 on extended regions. 

 The stacked spectrum (see error bars Fig.~\ref{fig:BPT}) unambiguously confirms the high level of  [OI] emission outside the main \ion{H}{II} regions. Thus, the high [OI]/H$\alpha$ line ratios are unlikely to be due to measurement errors.

\subsubsection{Low-density Diffuse Ionized Gas}

The regions with strong [OI]/H$\alpha$ ratios surround the main star--forming knots; they lie within the warm low-density ionized phase of the ISM, often referred to as the Diffuse Ionized Gas (DIG). The DIG is ionized by field OB stars, leaking photons from H II regions \citep{Hoopes03}, or shocks. The low density of the DIG is known to harden the ionizing spectra and can lead to LINER-like line ratios. However, a very low ionization parameter, below log(U) $ < -4$ \citep[see][]{Hoopes03}, is required to reach the observed values of the [OI]/H$\alpha$ line ratios. 
As shown in Sect.~\ref{coeff_ion} , the data points on the [OI]/H$\alpha$ diagram constrain the ionization parameter to be in the range  $-3.5$ < log(U) < $-2$: this is much higher than the value needed to get non-stellar ionization-like line ratios in the DIG.

\subsubsection{Shocks}
\label{Shocks}
Shocks due to stellar winds, supernovae explosions, or dynamical processes such as collisions or accretion of gas can also produce large forbidden to Balmer line ratios.
We have tested the shock hypothesis with the fast radiative shock model from \citet{Allen08}, which is based on MAPPINGS III \citep{Sutherland93}.

\begin{figure*}
\centering
\includegraphics[width=18cm]{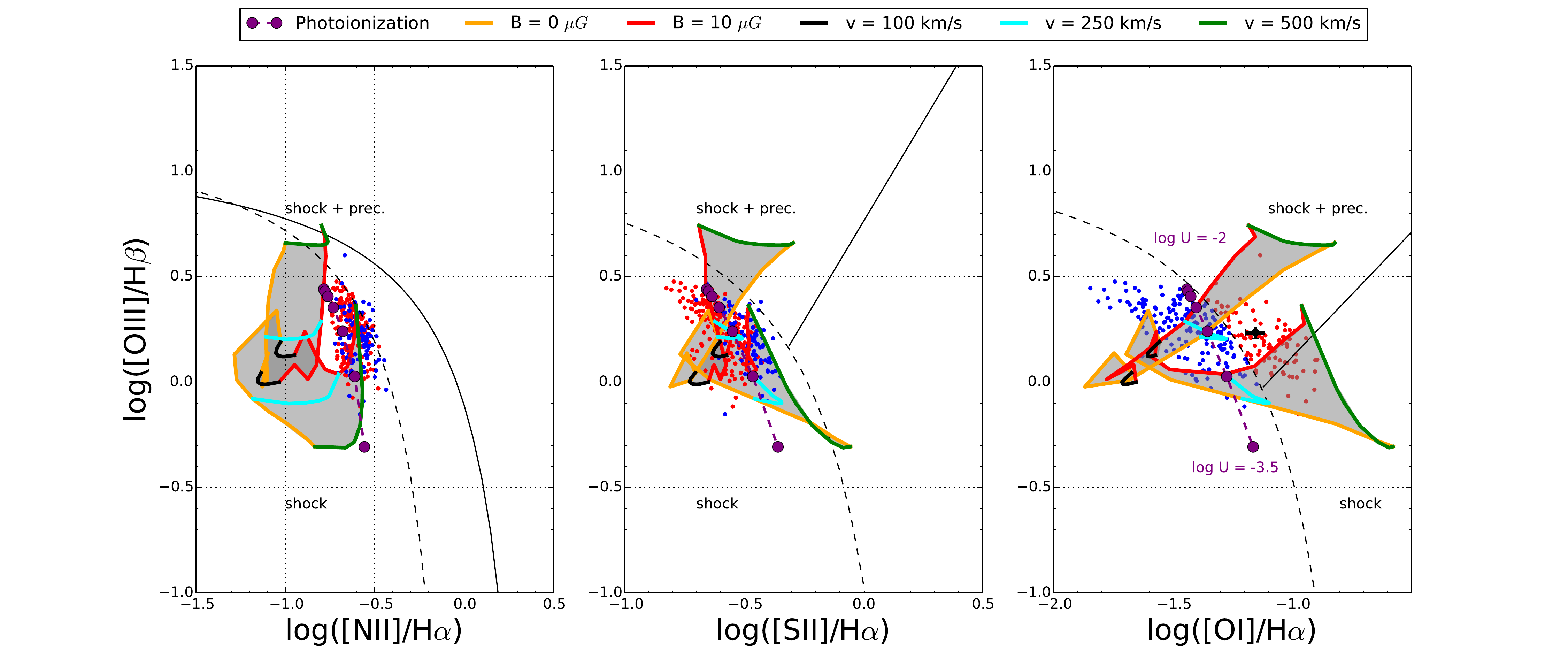}
\caption{Shock models on the BPT diagrams. Are plotted only points with detected [OI] emission. The blue (resp. red) points lie in (resp. out of) the star formation locus on the [OI] / H$\alpha$ BPT diagram. The purple dashed line is the ionization model prediction from MAPPINGS V, using a metallicity Z = 0.5 Z$_\sun$ and log(P/k) = 5.2. The ionization parameter goes from log U = -2 to log U = -3.5, as indicated on the third diagram. The two grids come from the MAPPINGS III model for fast-shock, as described in the text, and are drawn for shock only and shock + precursor. The locus of both grids are shaded.\label{shock}}
\end{figure*}

The shock model library has a limited number of input parameters.
We have assumed an electronic density of 1~cm$^{-3}$ and a LMC abundance, as these parameters are the closest available to that estimated in the ISM of NGC~5291N.
The speed of the shock and the transverse magnetic parameter were varied, from 100 to 300~km~s$^{-1}$ and from 0 to 10~$\mu$G respectively.
The locus of the resulting model is shown in Fig.~\ref{shock} for shock only and shock + precursor. We find that the outlying data points are encompassed between the grids for shock and shock+precursor. 
For comparison, the prediction for pure stellar photo-ionization from MAPPINGS V is also shown in Fig.~\ref{shock}. The average electronic density of the points lying outside of the 'star forming' region on the third BPT diagram being around 10~cm$^{-3}$, we use the model with log(P/k) = 5.2 (see Sect.~\ref{coeff_ion}). A half-solar metallicity was also assumed.

\begin{figure}[]
\includegraphics[width=9cm]{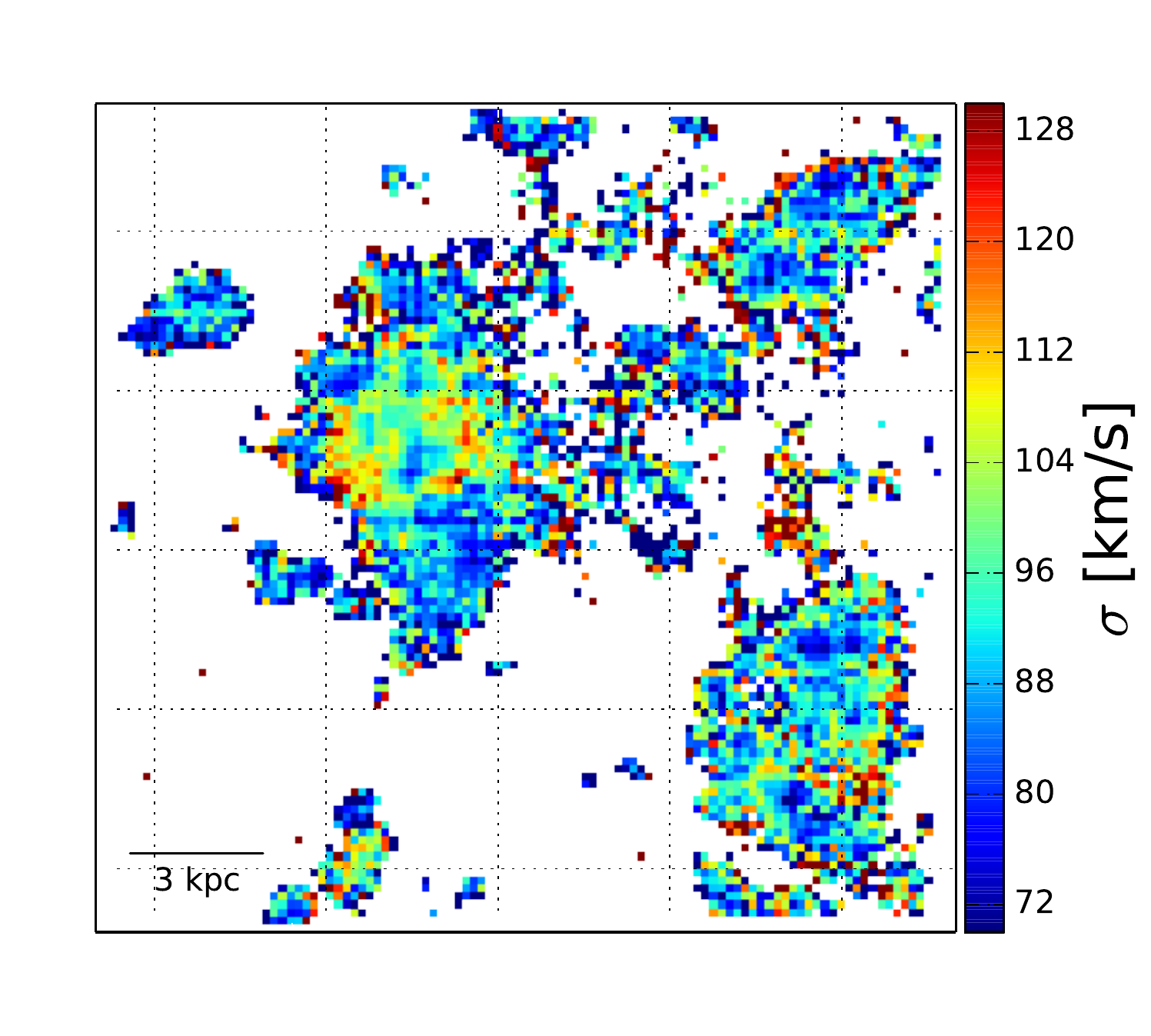}
\caption{Velocity dispersion measured from the width of the H$\alpha$ emission line. The field of view is the same as in Fig.~\ref{fig:att}.\label{veldisp}}
\end{figure}

 The velocity dispersion map of the ISM was computed from the width of the H$\alpha$ emission line. We subtracted a constant value of 2.34 $\AA$ to the width of the H$_{\alpha}$ line, corresponding to the average Gaussian FWHM close to the H$\alpha$ line. The map, shown in Fig.~\ref{veldisp}, has peaks at $\sigma$ > 130~km~s$^{-1}$ in the central regions -- most likely due to spatially unresolved rotation \citep{Bournaud04} -- but also in the outskirts of the main star forming knots close to the region where [OI]/H$\alpha$ culminates.
A detailed inspection reveals that the high-$\sigma$ regions are in fact located a bit further away where the [OI] emission is below the detection limit. However, the observed line-of-sight velocity dispersions could be consistent with transverse shocks of up to 300~km~s$^{-1}$. If shocks are responsible for the high [OI]/H$\alpha$ ratios, what is their origin?
 
First, a shock could have arisen from the accretion of gas onto the newly formed dwarf. In this case we would expect the shocks to be present only around the galaxy because the accretion of gas is supposed to happen towards the deepest gravitational potential well, hence all around the dwarf galaxy. However, as one can see in Fig.~\ref{fig:BPT}, the strong [OI]/H$\alpha$ ratio regions are not only found around the kinematically decoupled dwarf galaxy but around most star--forming regions away from NGC~5291N (See the South-West clump for instance).

Second, a shock could be due to the stellar feedback from the very recent starburst. Strong outflows in the centre of the star--forming region would then be expected. The expected kinematical signature of such outflows -- an additional broad component in the emission lines -- is not seen in the MUSE data: single Gaussian fits properly fit the H$\alpha$ emission line, even in the centre of the star--forming regions. 
\footnote{Fabry-Perot H$\alpha$ data with high spectral resolution for this region are available \citep{Bournaud04}. Their spectral resolution (R = 9375) are far superior to the MUSE data (R = 2840 close to H$\alpha$). Their analysis reveal complex kinematical features, with likely superimposed foreground and background clouds, but no obvious signature of outflows. The broad wings of the H$\alpha$ line in their data are instrumental signatures.}

Third, the high $\sigma$ velocities could be the remaining signature of the violent collision between NGC~5291 and a massive interloper that created the HI ring. 
The shock signature could have been wiped out in the central star--forming region because of the stellar activity and only remain in the low-density gas surrounding this area. In the simulation of \cite{Bournaud07} of the ring formation and expansion, the induced velocity dispersion 0.7 Gyr after the collision had decreased to about $\sigma \simeq$ 40~km~s$^{-1}$. According to the shock models, this is too low to account for the high [OI] / H$\alpha$ flux ratio.

Therefore, the models involving a photoionization within a very low density environment (the DIG hypothesis) or shocks do not provide satisfactory explanations for the values and spatial variations of the [OI]/H$\alpha$ flux ratios. Confirming the existence of regions with high [OI]/H$\alpha$ in the clumpy ISM of collisional debris (or of distant galaxies) should be a first step in our understanding of the ionization process.
In fact, \citet{Cairos10}, who obtained IFU data for eight dwarf galaxies considered as BCDGs with the Postdam Multi-Aperture Spectrophotometer (PMAS) at the 3.5~m telescope at Calar Alto Observatory, also measured high values of [OI]/H$\alpha$ -- as high as  log([OI]/H$\alpha$) = $-0.50$ -- for several objects in their sample. It is remarkable that these high ratios are always located in the outskirts of the galaxies. In some of them, namely I~Zw~159 and Mrk~32, they also suggest the presence of shocks in the same regions, based on strong values of the [SII]/H$\alpha$ ratio, above log([SII]/H$\alpha$) = $-0.20$ . However the lack of spatial resolution prevents firm conclusions. 
Strong outflows presumably generating shocks have been found in ESO338-IG04, the only other dwarf galaxy for which MUSE data are available \citep{Bik15}.
Outflows were disclosed in the analysis of the H$\alpha$ kinematic maps and highly ionized cones were probed by using the [SII]/[OIII] line ratio, thought to be escape gates to Ly$\alpha$ photons. It is interesting to note that no such feature was observed in NGC~5291N.

\subsection{Origin of the diffuse continuum emission\label{sect:emission_discussion}}
\begin{figure*}[]
\includegraphics[width=19cm]{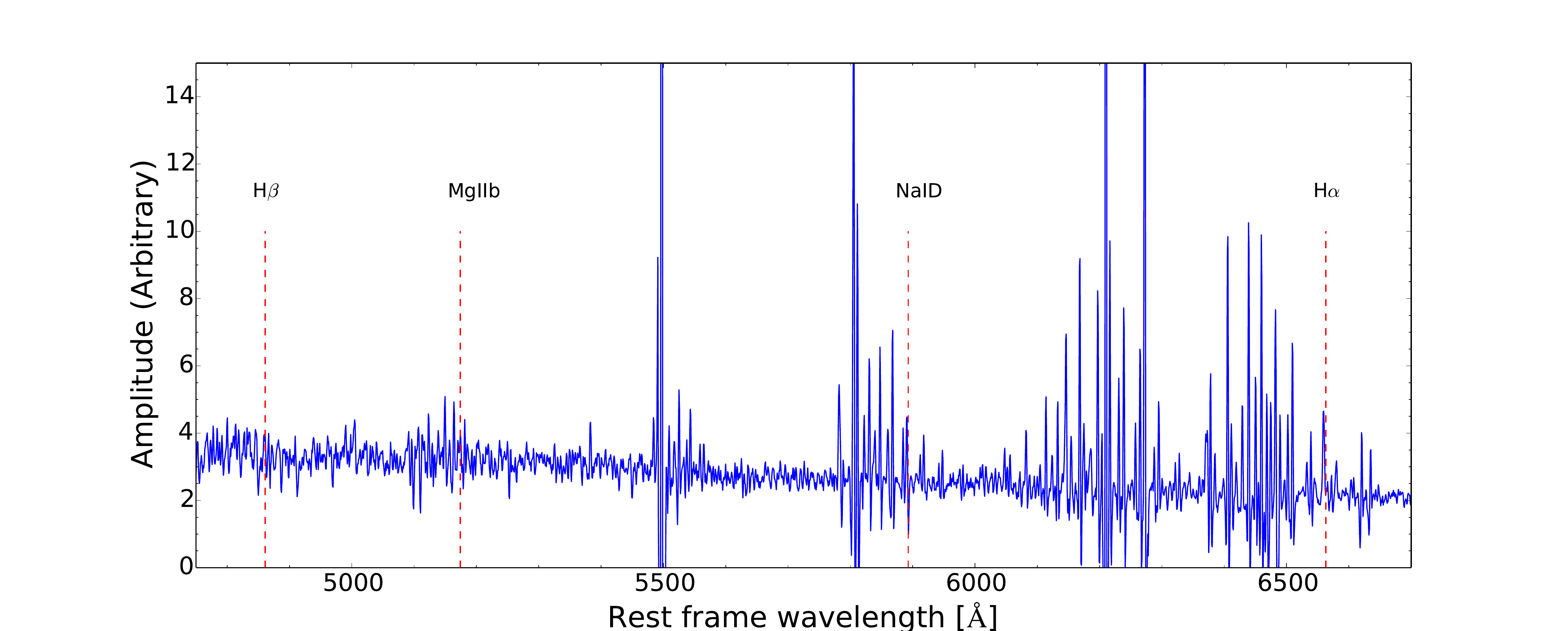}
\caption{Stacked spectra from the South-East region indicated by a black rectangle on Fig.~\ref{fig:cont}. Every spaxel showing a 3$\sigma$ H$\alpha$ detection was rejected.\label{cont_spectra}}

\end{figure*}

As presented in Sect.~\ref{sect:emission}, the H$\alpha$ emitting regions are surrounded by a diffuse blue light emission of still unknown origin.
One may speculate that this light has been emitted by a stellar population old enough to no longer ionize the surrounding gas. This would be consistent with the outside-in evidence of the star--formation events derived from the H$\beta$ equivalent width map.
To further characterize the physical properties of this diffuse component, we have stacked the spectra over the full South-East region where the diffuse continuum emission is most prominent. The resulting spectrum does not reveal any underlying absorption line (See Fig.~\ref{cont_spectra}), nor actually residual emission lines. This puts very strong constraints on the age of the putative underlying stellar population; galaxy SED models such as GALEV \citep{Kotulla09} show that the transition between emission to absorption line for H$\beta$ occurs at around 25~Myr after an instantaneous burst, and does not last more than a few Myr. One would then witness a special episode of the star formation history: the transition between the death of the OB stars and emergence of type A absorption lines, lasting a very short time but occurring on an extended region, spanning over 3 kpc. This would mean that the star--formation episode occurred almost simultaneously on a 3 kpc scale region, which is rather unrealistic.

\begin{figure}[]
\includegraphics[width=9cm]{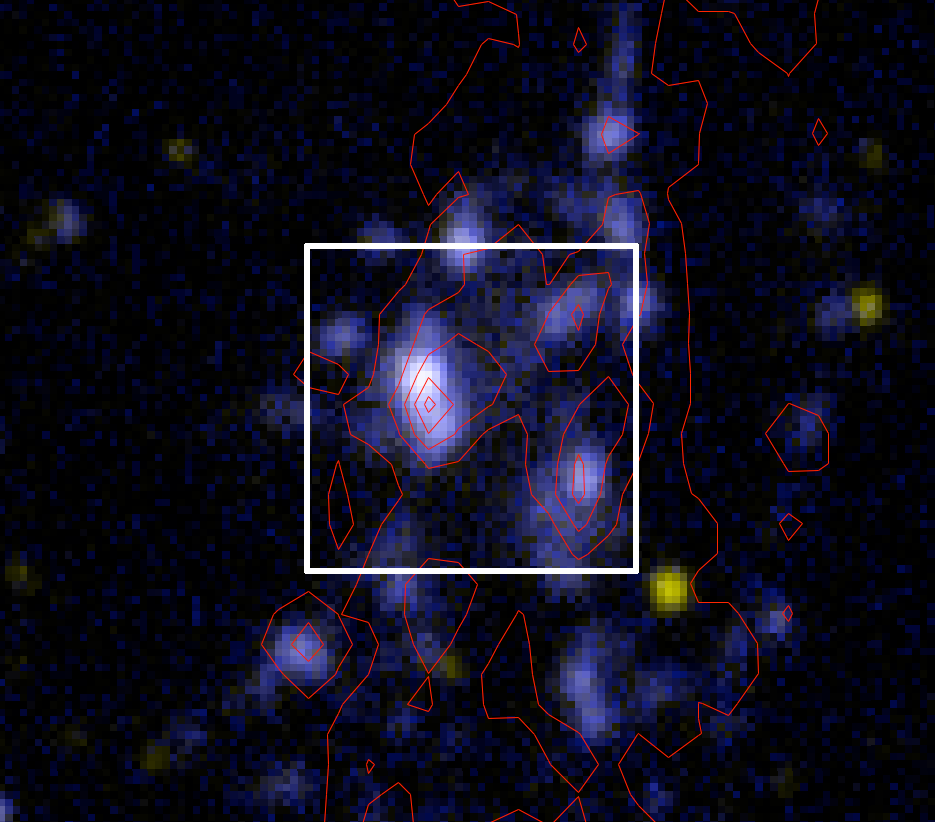}
\caption{Composite color image of the UV emission as seen by GALEX: the near-UV is shown in red and green and the far-UV in blue. HI contours are superimposed in red. The MUSE field of view is shown by the white square.\label{UV}} 
\end{figure}

A second hypothesis would be that this emission is actually scattered light from the star--forming region. In this case, one should expect to see an associated UV counterpart, that is dust scattered leaking UV emission from the hot young stars \citep[see][and references therein]{Boissier15}. 
However GALEX data \citep[see description in][]{Boquien07, Boquien09} of this system showed that there is no near-UV nor far-UV emission towards the optical blue diffuse component (compare Fig.~\ref{fig:cont} and Fig.~\ref{UV}). One should note that the GALEX data are quite deep, with exposition time of about 9000~s for the NUV and 5500~s in FUV.

Thus, the origin of this diffuse component remains undetermined. MUSE observations of other similar systems (or of other parts of the NGC~5291 gaseous ring) should tell how common it is around starbursting galaxies.

\section{Conclusions}
\label{conclusion}

We have presented FORS deep multi-band optical images and high spatial resolution IFU data on NGC~5291N, one of the very few starbursting dwarfs yet observed with the revolutionary instrument MUSE on the ESO VLT.
 This dwarf, located at 63~Mpc, formed within a ring of gas expelled from the host galaxy, NGC~5291, after a violent collision several 10$^{8}$ years ago. NGC~5291N shares a number of physical properties with clumps in gas rich star forming high redshift galaxies: it is very gaseous, has a clumpy ISM with sub-solar metallicity and contains no old stars, contrary to classical blue compact dwarf galaxies.
 
The MUSE data cube provided us with optical spectra, in the range 4750 - 9350~$\AA$ over a field of view of about 1\arcmin$\times$1\arcmin. Each spaxel, corresponds to 60~pc~$\times$~60~pc at the distance of NGC~5291N , before re-binning the data-cube. Strong emission lines (H$\beta$, [OIII], [OI], H$\alpha$, [NII] and [SII]) and continuum emission were extracted and mapped to investigate the properties of the ISM. The derived maps showed a rather strong heterogeneity in the spatial distributions of the different emission lines and an extended blue diffuse continuum emission. 

Emission line ratios were investigated with BPT-like diagnostic diagrams. They showed that most data point are consistent with star--forming regions, with a noticeably large scatter on the diagrams. Furthermore, a significative number of points fall outside of the star--formation locus of the [OIII]/H$\beta$ vs [OI] / H$\alpha$ diagnostic diagram. They correspond to regions located on the outskirts of the dwarf galaxy.

The equivalent width of the H$\beta$ (EW(H$\beta$)) line shows that the galaxy has undergone a very recent starburst: no older than 3~Myr in the brightest region to 7~Myr old in the fainter neighborhood, under the hypothesis of an instantaneous burst. The observed distribution of EW(H$\beta$) is consistent with an outside-in star formation episode.
The extinction map (Av) computed from the Balmer line ratio H$\alpha$ / H$\beta$ indicates that the dust concentrates towards the most active star--forming regions, where mid and far infrared emission had already been found. This is also the region where the V-R color map derived from the FORS broad-band images shows narrow red filaments that are hence most likely associated with dust lanes. 

The metallicity was computed using the empirical O3N2 calibration and gave a mean value <12 + log(O/H)> = 8.4 with a standard deviation of 0.05 (equivalent to an abundance of Z~$\simeq$~0.5~Z$_{\odot}$), confirming previous estimates from slit spectroscopy. This relatively high value is fully expected for a recycled dwarf galaxy made in collisional debris. A surprising  inside-out positive metallicity gradient is observed; it is at odds with predictions from numerical simulations of the chemical evolution of Tidal Dwarf Galaxies. However, the metallicity distribution spreads over less than 0.2~dex, which is the intrinsic uncertainty of the empirical method. The computed O3N2 map is therefore consistent with an homogeneous metallicity coupled with variations of other physical parameters, such as the ionization parameter. 

The spatial distribution of the ionization parameter was determined through a [OIII]/[SII] versus [NII]/[SII] photoionization grid computed with the MAPPINGS V model. The inferred ionization parameter spreads over a large range of values, from log(U) = -3.5 to log(U) > -2, explaining the observed scatter on the BPT diagrams, except for the points with high [OI] / H$\alpha$ flux ratios which are not compatible with simple stellar photoionization models.
Models involving the low-density DIG which is hot, only partially ionized, can emit non-stellar emission line ratios, but they need a ionization parameter lower than the one derived from the MUSE data.

Various shock models were investigated. They can reproduce the observed flux ratios not compatible with stellar photoionization for shock velocities higher than 200~km~s$^{-1}$. 
The velocity dispersion map computed from the IFU data cube do show an increase of $\sigma$(V) outside  the main star--forming regions, but at a level and precise location which are not fully consistent with the shock model. 

An extended blue diffuse emission surrounding the main star--forming regions is observed in the deep optical FORS images and detected with MUSE. The IFU data indicates that this is pure continuum emission, exhibiting no emission nor absorption lines, expanding over distances of at least 3~kpc, from the emission line regions. The hypothesis of emission arising from a stellar population old enough not to ionize the surrounding HI gas, but also young enough not to exhibit Balmer absorption lines, appears rather unlikely. Besides, the UV emission associated with scattered light from the central star formation region was not observed by the GALEX UV observatory. 

Therefore no firm conclusions on the origin of the deviant points in the [OI] / H$\alpha$ BPT diagram, spatially located just outside the starburst regions of the dwarf, and of the pure diffuse blue optical continuum emission found further out, could be reached. This calls for further IFU observations of the outskirts of nearby star--forming galaxies, in particular the youngest ones formed in debris of galaxy-galaxy collisions, arguably the best local proxies of the distant star--forming objects. 
  
\begin{acknowledgements}

The authors thank Michael Dopita and Stephanie Juneau for very useful discussions, and the (anonymous) referee for in-depth reading and helpful comments. P.M.W received funding through BMBF Verbundforschung (project MUSE-AO, grant 05A14BAC). E.Z acknowledges funding from the Swedish Research Council (project 2011-5349). This research has made use of the NASA/IPAC Extragalactic Database (NED) which is operated by the Jet Propulsion Laboratory, California Institute of Technology, under contract with the National Aeronautics and Space Administration.

\end{acknowledgements}

\bibliographystyle{aa}
\bibliography{draft}

\end{document}